\documentclass[onecolumn, draftclsnofoot,11pt]{IEEEtran}
\IEEEoverridecommandlockouts
\usepackage{cite}
\usepackage{amsmath,amssymb,amsfonts}
\usepackage{algorithmic}
\usepackage{graphicx,color}
\usepackage{textcomp}
\usepackage{bm}

\newtheorem{Theorem}{\sc Theorem}
\newtheorem{Lemma}{\sc Lemma}
\newtheorem{Proposition}{\sc Proposition}
\newtheorem{Corollary}{\sc Corollary}

\newenvironment{Proof}{\par  \sc
	Proof.\rm}{\hspace*{\fill}$\square$\vspace{1ex}}

\def\BbbZ{{\sf Z\hspace*{-0.95ex}Z}}
\newcommand{\Z}[1]{\relax\ifmmode\BbbZ_{#1}\else $\BbbZ_{#1}$\fi}

\newcommand{\PP}{\mathcal P}

\newcommand{\cancel}[1]{}

\newcommand{\A}{\mathcal{A}}

\def\BibTeX{{\rm B\kern-.05em{\sc i\kern-.025em b}\kern-.08em
    T\kern-.1667em\lower.7ex\hbox{E}\kern-.125emX}}

\begin{document}
\bibliographystyle{IEEEtran}

\title{Algebraic Geometric Secret 	Sharing Schemes over Large Fields Are Asymptotically Threshold}

\author{Fan Peng, Hao Chen and Chang-An Zhao
	\thanks{The research  of Chang-An Zhao was supported by National Key R$\&$D Program of China under Grant 2017YFB0802500.  The research of Hao Chen was supported by NSFC Grants 11531002, 62032009 and the Major Program of Guangdong Basic and Applied Research Grant 2019B030302008. The research  of Chang-An Zhao was also partially supported by NSFC Grant 61972428, the Major Program of Guangdong Basic and Applied Research under Grant 2019B030302008 and the Open Fund of State Key Laboratory of Information Security (Institute of Information Engineering, Chinese Academy of Sciences, Beijing 100093) Grant 2020-ZD-02.	}

\thanks{F. Peng is with College of Mathematics and Statistics, Guangxi Normal
	University, Guilin, China.
	{(E-mail: pengfan@gxnu.edu.cn}.}

\thanks{H. Chen is with  College of Information Science and Technology$/$College of
	Cyber Security, Jinan University, Guangzhou, Guangdong  Province,
	510632, China.
	{E-mail: haochen@jnu.edu.cn}}

\thanks{C.-A, Zhao is with  School
	of Mathematics, Sun Yat-sen University, Guangzhou 510275, P.R.China and with 	Guangdong Key Laboratory of Information Security,	Guangzhou {\rm 510006}, P.R. China.
	(E-mail: zhaochan3@mail.sysu.edu.cn)}
}

\maketitle

\begin{abstract}
In Chen-Cramer Crypto 2006 paper \cite{cc} algebraic geometric secret
sharing schemes were proposed such that the ``Fundamental Theorem in
Information-Theoretically Secure Multiparty Computation" by Ben-Or,
Goldwasser and Wigderson \cite{BGW88} and Chaum, Cr\'{e}peau and
Damg{\aa}rd \cite{CCD88} can be established over constant-size base
finite fields. These algebraic geometric secret sharing schemes defined
by a curve of genus $g$ over a constant size finite field ${\bf F}_q$
is quasi-threshold in the following sense, any subset of $u \leq T-1$ players (non qualified)
has no information of the secret and any subset of $u \geq T+2g$ players
(qualified) can reconstruct the secret. It is natural to ask that how far
from the threshold these quasi-threshold secret sharing schemes are?
How many subsets of $u \in [T, T+2g-1]$ players can recover
the secret or have no information of the secret?

In this paper it is proved that almost all subsets of $u \in
[T,T+g-1]$ players have no information of the secret and almost all
subsets of $u \in [T+g,T+2g-1]$ players can reconstruct the secret
when the size $q$ goes to the infinity and the genus satisfies
$\lim \frac{g}{\sqrt{q}}=0$. Then algebraic geometric secret
sharing schemes over large finite fields are asymptotically
threshold in this case. We also analyze the case when the size
$q$ of the base field is fixed and the genus goes to the infinity.\\
\end{abstract}


\section {Introduction}

\subsection{Linear secret sharing schemes (LSSS) and applications}

Let $K$ be a finite field. A {\em $K$-linear secret sharing scheme}
(LSSS)$\{K,V_1,...,V_m,u\}$ on a set of participants
$\mathcal{P}=\{P_1,...,P_m\}$ is defined as a sequence of subspaces
$\{V_1,...,V_m\}$ of $K^e$, where $V_i \subset K^e$ and $u$ is a
given vector in $K^e$. A subset $A$ of $\mathcal{P}$ is qualified if
$u$ is in the subspace of $K^e$ spanned by the $\{V_i\}_{i \in A}$.
For any secret value $s \in K$, choose a random linear mapping $\phi
: K^e \rightarrow K$ such that $\phi(u)=x$,
$\{\phi(e_1^i),...,\phi(e_{dimV_i}^i)\}$ is the share of the
participant $P_i$, where $e_1^i,...,e_{dimV_i}^i$ is a base of $V_i
\subset K^e$. Only the qualified subsets of $\mathcal{P}$ can
reconstruct the secret from their shares. The {\em access
	structure}, $\Gamma \subset 2^{\mathcal{P}}$, of a secret-sharing scheme
is the family of all qualified subsets of $\mathcal{P}$. The {\em
	adversary structure} $\Gamma^c$ is the family consisting of all
subsets of $\mathcal{P}$ not in $\Gamma$. The minimum access
structure $\min \Gamma \subset 2^{\mathcal{P}}$ is defined to be the
set of all minimum elements in $\Gamma$ (here we use the natural
order relation $ S_1 <S_2$ if and only if $S_1 \subset S_2$ on
$2^{\mathcal{P}}$). We call a secret-sharing scheme a {\em
	$(k,m)$-threshold scheme} if the access structure consists of all
the subsets of $\mathcal{P}$ with $k$ or more elements, where the
cardinality of $\mathcal{P}$ is $m$. The first secrets-sharing
scheme was proposed independently by Shamir \cite{sha} and Blakley
\cite{bl} in 1979, and is in fact a threshold secret-sharing scheme.
The existence of secret-sharing schemes with arbitrary given access
structures was proved in \cite{BL}. The complexity of the $K$-LSSS is
defined to be $\lambda(\Gamma)=\Sigma_{i=1}^m \dim_K(V_i)$. When the
complexity is $m$, the LSSS is called {\em ideal}. One of the main
open problems in secrete sharing is the characterization of the
access structures of ideal secret sharing schemes \cite{bri}.\\

Let $\{K,V_1,...,V_m,u\}$ be an LSSS over $K$, denote $v=u \otimes u
\in K^{e^2}$ and $V_i'=V_i \otimes V_i \in K^{e^2}$  for
$i=1,...,m$. This LSSS is said to have multiplicative property if
$v$ is in the linear subspace of $K^{e^2}$ spanned by all
$\{V_i'\}_{i=1,...,m}$. This is equivalent to the following fact :
for any given two secrets $x$ (with shares $c_i \in K^{dimV_i}$ for
$i=1,...m$) and $y$(with shares $d_i \in K^{dimV_i}$ for
$i=1,...,m$), the product $xy$ is in the linear span with coefficients
in $K$ of $c_id_i$ (here the product can be understood as a bilinear
mapping $K^{dimV_i} \otimes K^{dimV_i} \rightarrow K^{dimV_i}$). It
is said to have strongly multiplicative property if the $v$ is in
the linear subspace spanned by $\{V_i'\}_{i \in {\bf P}-B}$ where
$B$ is any subset of $\mathcal{P}$ in the adversary structure.\\

For an adversary structure $\Gamma^c$ on $\mathcal{P}$, it is said
that $\Gamma^c$ is $Q_2$ if $A\bigcup B\neq \mathcal{P}$ for any
$A,B \in \Gamma^c$, and $\Gamma^c$ is $Q_3$ if $A\bigcup B \bigcup
C\neq \mathcal{P}$ for any $A,B,C \in \Gamma^c$. One of the key
results in \cite{CDM} is a method to construct, from any LSSS with a
$Q_2$ access structure $\Gamma$, a multiplicative LSSS $\Gamma'$ with the same
access structure and double complexity, that is $\lambda(\Gamma') \leq
2\lambda (\Gamma)$ can be constructed. $K$-MLSSSs with $Q_2$ and $Q_3$ access
structures are closely related to secure multi-party computation. It
is known that any strongly multiplicative LSSS can be efficiently
transformed into a polynomial complexity error-free multi-party
computation protocol computing any arithmetic circuit. This protocol
is information-theoretically secure against the adaptive and active
$\Gamma^c$ adversary. For details of secure multi-party computation
and its relation with linear secret sharing schemes, the reader is
referred to \cite{CDM,cdn}.\\

The approach of secret sharing based on error-correcting codes was
studied in \cite{ma1,ma2,mcsa,CCGdeV}. It is a special form of the above
LSSS. Actually it was realized that Shamir's
$(k,n)$-threshold scheme is just the secret sharing scheme based on
the famous Reed-Solomon code in 1979 paper \cite{mcsa}. \\

We recall the construction of LSSS from
error-correcting codes in \cite{CCGdeV}. Let $\textbf{C}$ be a $q$-ary
$[n + 1,k,d]$-code, let $G=(\textbf{g}_0 ,\textbf{g}_1, \cdots,
\textbf{g}_n)$ be a generator matrix for $\textbf{C}.$  We give a
construction for a secret sharing scheme for $\mathcal{P}=\{P_1,
\cdots, P_n\}$ as follows:

(1) Let the generator matrix $G$ be publicly known to everyone in
the system.

(2) To share a secret $s\in \mathbb{F}_{q},$ the dealer randomly
selects a vector $$\textbf{r}=(r_1 ,r_2,\cdots, r_k)\in
\mathbb{F}_{q}^k$$ such that $s=\textbf{r}\cdot \textbf{g}_0.$

(3) Each participant $P_i$ receives a share $s_i=\textbf{r}\cdot
\textbf{g}_i,$ for $i=1, \cdots, n.$

We have the codeword $\textbf{c}=(s, s_1, \cdots, s_n)=\textbf{r}G.$
This is an ideal perfect secret sharing scheme. We refer to
\cite{ma1,ma2} and\cite{mcsa} for the following Lemma.\\

\begin{Lemma}
	Let $\textbf{C}$ be a linear code of length $(n+1)$ with generator
	matrix $G$. Suppose the dual of $\textbf{C},$ i.e.,
	$\textbf{C}^{\bot}=\{\textbf{v}= (v_0, v_1, \cdots, v_n)|
	G\textbf{v}^T=0\},$ has no codeword of Hamming weight $1$. In the
	above secret sharing scheme based on the error-correcting code
	$\textbf{C},$ for any positive integer $m,$ $\{P_{i_1}, \cdots,
	P_{i_m}\}$ can reconstruct the secret if and only if there is a
	codeword $\textbf{v} =(1, 0, \cdots, v_{i_1}, \cdots, v_{i_m},
	\cdots, 0)\in \textbf{C}^{\bot},$ i.e., the support of the codeword
	$\mathrm{supp}(\textbf{v})\subseteq \{0, i_1, \cdots, i_m\}.$
\end{Lemma}

\subsection{Algebraic geometric secret sharing schemes over a
	constant-size field}

Secret sharing schemes proposed in \cite{cc} can be thought as a
natural generalization of Shamir's scheme by applying
algebraic-geometric codes in the
above construction. In Shamir's scheme
the number of players has to be upper bounded by the size of the base
field. In algebraic geometric secret sharing schemes this
restriction can be removed with quasi-threshold access structures
instead of threshold access structures . These LSSS from algebraic
curves are quasi-threshold in the following sense, any subset of $u
\leq T-1$ players (non qualified) has no information of the secret
and any subset of $u \geq T+2g$ players (qualified) can reconstruct
the secret, where $g$ is the genus of the curve on which the secret
sharing is defined. Algebraic geometric secret sharing schemes have
a remarkable application in the "Fundamental Theorem in
Information-Theoretically Secure Multiparty Computation" by Ben-Or,
Goldwasser and Wigderson \cite{BGW88} and Chaum, Cr\'{e}peau and
Damg{\aa}rd \cite{CCD88}. The communication complexity in the  above
fundamental protocols is saved with a $\log(n)$ factor and the
information-theoretically secure multiparty computation can be
established over a constant-size field $\mathbb{F}_q$ with a
decreasing corruption tolerance by a small
$\frac{1}{\sqrt{q}-1}$-fraction (see \cite{Cramer,cdn}). The
asymptotical result in \cite{cc} plays a central role in
\cite{IKOS2007} about communication-efficient zero knowledge for
circuit satisfiability, two-party computation \cite{DIK,DZ,IPS}, OT
combiners \cite{HIKN} and correlation extractors \cite{IKOS2009}. We
refer to \cite{cdn} page 342 for its impact in secure multiparty
computation and other fields of
cryptography.\\

Let $\mathbb{F}_q$ be a given finite field with $q$
elements, $C$ be a smooth projective absolutely irreducible
curve defined over $\mathbb{F}_q$ with the genus $g$, and
$C(\mathbb{F}_q)$ be the set of ($\mathbb{F}_q$) rational points of
$C$. $\{Q, P_0, P_1,\cdots,P_n\}$  is a subset of
$C(\mathbb{F}_q),$  and $G=mQ.$ $D$ is the divisor $P_0+P_1+\cdots+P_n$. Let
$L(G)=\big\{f\in\mathcal{M}_C\big|\mathrm{div}(f)+G\geq0\big\}$ with
dimension denoted by $l(G),$ and let
$\Omega(G-D)=\big\{\omega\in\Omega_C\big| \mathrm{div}(\omega)\geq
G-D\big\}$ with dimension denoted by $i(G-D).$

We can define algebraic geometry codes
$$C_{L}(D,G)=\big\{(f(P_0), f(P_1),\cdots,f(P_n))\big|f\in L(G) \big\},$$
and
$$C_{\Omega}(D,G)=\big\{(\mathrm{res}_{P_0}
(\omega), \mathrm{res}_{P_1},
(\omega)\cdots,\mathrm{res}_{P_n}(\omega))\big|\omega\in
\Omega(G-D)\big\}.$$

Assume $2g-2<\mathrm{deg}G=m<n+1,$ then $l(G)=m-g+1$ and
$i(G-D)=n-m+g$ by the Riemann-Roch theorem. The first one is a linear
$[n+1, k, d]$ code, and the second one is the dual $[n+1, n+1-k,
d^{\bot}]$ code, where $k=m-g+1, d\geq n+1-m,$ and $d^{\bot}\geq
m-2g+2.$

Applying Massey's construction as follows.

(1) To share a secrete $s\in \mathbb{F}_q,$ the dealer randomly
select an element $\omega\in \Omega(G-D)$ such that
$\mathrm{res}_{P_0}(\omega)=s.$

(2) The share of the participant is
$s_i=\mathrm{res}_{P_i}(\omega)\in \mathbb{F}_q,$ for $1\leq i\leq
n.$

Note that, in (1) of the construction above, due to Riemann-Roch
theorem, we have $i(G-D)-i(G-D+P_0)=1.$ Thus there exists a non-zero
rational differential $\omega\in
\Omega(G-D)\setminus\Omega(G-D+P_0)$ such that
$\mathrm{res}_{P_0}(\omega)=s.$ Then we have a codeword $(s, s_1,
\cdots, s_n)\in C_{\Omega}(D,G).$ Let $\mathcal{A}=\{P_{i_1},
\cdots, P_{i_t}\}\subset \mathcal{P}$ and
$P_{\mathcal{A}}=\sum_{P\in \mathcal{A}}P,$ then the complement
$\mathcal{A}^c=\mathcal{P}\setminus\mathcal{A}$ is qualified if and
only if there exists a rational function $f\in L(G)$ such that
$f(P_0)\neq 0,$ and $f(P)=0,$ for $P\in \mathcal{A},$ i.e. $f\in
L(G-P_{\mathcal{A}})$ with $f(P_0)\neq 0.$ So we have following
three cases.

(1) When $t\leq m-2g,$ since
$\mathrm{deg}(G-P_{\mathcal{A}})\geq2g,$ so the linear system
$L(G-P_{\mathcal{A}})$ is base point free \cite{hart}. Then there is
a function $f$ with the above property, thus $\mathcal{A}^c$ is
qualified.

(2) When $t>m,$ since $\mathrm{deg}(G-P_{\mathcal{A}})<0,$ so
$L(G-P_{\mathcal{A}})=\emptyset,$ thus $\mathcal{A}^c$ is
unqualified.

(3) When $m-2g<t\leq m,$ it's hard to determine $\mathcal{A}^c$ is
qualified or unqualified.\\

In \cite{clx} access structures of LSSS from elliptic curves were
completely determined.

\begin{Theorem}[Chen, Lin, Xing]\label{CLX}
	Let $E$ be an elliptic curve over $\mathbb{F}_q,$ and let $\{P_0,
	\cdots,  P_n\}$ be a subset of $E(\mathbb{F}_q)$ of $n + 1$ nonzero
	elements. Let $D=P_0+\cdots +P_n$ and $G=mO.$ Consider the elliptic
	secret sharing scheme obtained from $E$ with the set of players
	$\mathcal{P}=\{P_1, \cdots, P_n\}$.
	
	Let $\mathcal{A}=\{P_{i_1}, \cdots, P_{i_t}\}$ be a subset of
	$\mathcal{P}$ with $t$ elements, and let $B$ be the element in
	$E(\mathbb{F}_q)$ such that the group sum of $B$ and $\{P_{i_1},
	\cdots, P_{i_t}\}$ in $E(\mathbb{F}_q)$ is $O$. If
	$\mathcal{A}^c\stackrel{\mathrm{def}}{=}\mathcal{P}\setminus
	\mathcal{A}$ is a minimal qualified subset for the secret sharing
	scheme from $C_{\Omega}(D, G)$, then $t\leq m$ Furthermore, we have
	the following:
	
	1) when $t=m, \mathcal{A}^c$ is a minimal qualified subset if and
	only if $B=O.$
	
	2) when $t=m-1, \mathcal{A}^c$ is a minimal qualified subset if and
	only if $B\notin \{P_0, \cdots,  P_n\}$ or $B\in\mathcal{A}$.
	
	3) any subset of $\mathcal{P}$ of more than $n-m+2$ elements is
	qualified.
\end{Theorem}

\section{Main result and open question}

\subsection{Main result}

In this paper  we will discuss the access structures of algebraic geometric secret
sharing schemes when $q$ tends to the infinity and the genus satisfies
$\lim \frac{g}{\sqrt{q}}=0$. Roughly speaking
quasi-threshold algebraic geometric schemes approach to the
threshold secret sharing in this case.\\

The access structures of elliptic curve secret sharing schemes are
determined completely in \cite{clx}, by
applying the finite Abelian group structure of $E(\mathbb{F}_q).$ We
will analyze elliptic curves and higher genus curves cases when the order of the ground field $q$ tends to the infinity. In the elliptic curve case, because  the set of  the
rational points $E(\mathbb{F}_q)$ forms a finite Abelian group, we
reduce the problem to the counting of the number $N(t,B,\mathcal{P})$ of the set $$\left\{\mathcal{A}\in {\mathcal{P}\choose t}
\bigg|\mathop{\oplus}\limits_{P\in \mathcal{A}}P=B \right\},$$ where $\mathcal{P}\subset
\mathfrak{G}$ is a subset in $N$ element Abelian group $(\mathfrak{G},\oplus)$ with the cardinality $n$, ${\mathcal{P}\choose t}$ is the set of all subset of $\mathcal{P}$ of the cardinality $t$, and $B$ is an arbitrary element in $\mathfrak{G}$.\\

This problem has been
extensively studied in \cite {lw,lw1,lwz}. By Lemma (\ref{bd}),
the asymptotic formula $N(t,B,\mathcal{P})$ has a main term
${{n \choose t}}/{N}$ and an error term ${M\choose t}.$  Under our assumptions, we have $M<\delta'n,$ where $\delta'<1,$ so the error
term is much smaller than the main term, as $q \rightarrow \infty,$ then our result follows.\\

The same method is used for higher genus curve case with a
more complicated technique. The finite Abelian group is replaced by
the Jacobian variety $\mathrm{Jac}(C)(\mathbb{F}_q)$ over $\mathbb{F}_q$.
In Theorem 4 (or the main result below) by the using of the Abel-Jacobi map we express
the proportion of the qualified subsets by ${\sum\limits_{a\in
		\ominus W_{m-t}}N(t,a,\mathcal{P})}\Big/{{n \choose t}}.$  Further in
$(\star),$ it can be bounded by $$\frac{\big|\ominus
	W_{m-t}(\mathbb{F}_q)\big|}{{n \choose t}}\left\{\frac{{n \choose
		t}}{h_q(C)}+{M \choose t}\right\},$$ where $W_d=\phi_d(\mathrm{Sym}^d
C)$ is the image of symmetric product of the curve under the $d$-th Abelian-Jacobi mapping $\phi_d$ in $\mathrm{Jac}(C)$
and $h_q(C)=|\mathrm{Jac}(C)(\mathbb{F}_q)|$. By Weil bounds for character sums, we also have $M<\delta'n,$ where $\delta'<1.$
$h_q(C)$ have the classical Hasse-Weil
bound, and $\big|\ominus W_{m-t}(\mathbb{F}_q)\big|$ can be bounded
by Proposition 1. When $q \rightarrow \infty,$
all these bounds fit together to obtain our result.\\

Let $\mathbb{F}_q$ be a finite field with $q$
elements, $C$ be a smooth projective absolutely irreducible
curve defined over $\mathbb{F}_q$ with the genus $g$, and
$C(\mathbb{F}_q)$ be the set of ($\mathbb{F}_q$) rational points of
$C$. $\{Q,P_0,P_1,\cdots,P_n\}$ is a subset of
$C(\mathbb{F}_q)$, $D=P_0+P_1+\cdots+P_n$, and $G=mQ.$ for some $Q$ not in $supp(D)$. The set of players is $\mathcal{P}=\{P_1,\ldots,P_n\}$.
The following is our main result in this paper.\\

{\bf Main Result.} {\em We assume that $q \rightarrow \infty $ and
	$\lim\limits_{q\rightarrow \infty}\frac{g}{\sqrt{q}}=0.$ Suppose
	that $m=\delta n,$ where $\delta$ is a constant between $0$ and $\frac{2}{3},$
	$m$ and $n$ go to the infinity as $q$ tends to infinity. Suppose
	that $\big|C(\mathbb{F}_q)\big|-\big|\mathcal{P}\big|$ is bounded by
	a
	constant $c$ as $q$ tends to infinity.  Then
	\begin{itemize}
		\item[I)]
		when $0\leq m-t<g,$ the proportion of the qualified subsets
		approaches to zero;
		\item[II)]
		when $g\leq m-t<2g,$ the proportion of the qualified subsets
		approaches to~1.
\end{itemize}}.\\

\subsection{Open question}

When the size of the base field is fixed and the genus goes to the
infinity the situation would be quite different. In the above proof
of Theorem 4, the range of cardinalities of unknown subsets is $[T,
T+2g-1].$ The size of this range is $2g.$ Because $\lim
\frac{g}{\sqrt{q}}=0,$ and $n\sim \big|C(\mathbb{F}_q)\big|\sim q,$
then $\lim \frac{2g}{n}=0.$ On the other hand, if we fixed $q,$ and consider a maximal tower $\mathcal{C}=\{C_i\}$ over
$\mathbb{F}_q,$ which means that $\lim\limits_{i\rightarrow \infty}g(C_i)=\infty$
and $\lim\limits_{i\rightarrow \infty}\frac{|C_i(\mathbb{F}_q)|}{g(C_i)}=\sqrt{q}-1.$ When
$q$ is a square, this can be achieved \cite{GS}.
Because $\big|C_i(\mathbb{F}_q)\big|-\big|\mathcal{P}\big|<c,$ so the limit is $\lim \frac{2g}{n}
\approx\frac{2}{\sqrt{q}-1},$ In this case, the Weil bound $\Phi(\mathcal{P}) \leq(2g-2)\sqrt{q}+c\sim(2\frac{n}{\sqrt{q}-1}-2)\sqrt{q}\sim2\frac{\sqrt{q}}{\sqrt{q}-1}n$ is weaker than the trivial bound $\Phi(\mathcal{P}) \leq n.$ By this trivial bound, we have $M>n,$ so ${n \choose t}<{M \choose t}.$ The estimate about $(\star)$ fails. We
refer to \cite{Cascudo} for some results on the threshold gap of
secret sharing. It seems that in the case $q$ is fixed and the genus
goes to the infinity algebraic secret sharing schemes over a fixed
base field are not
asymptotically threshold.\\

\section{Technical tools}

In this section for $x\in \mathbb{R},$  we denote $(x)_0=1$ and
$(x)_t=x(x-1)\cdots(x-t+1)$ for $t\in \mathbb{Z}^+.$ For $t\in
\mathbb{N}, {x \choose
	t}\stackrel{\mathrm{def}}{=}\frac{(x)_t}{t_!}.$

\subsection{Li-Wan's sieve}

We recall the sieving formula discovered by Li and Wan
\cite{lw}. Roughly speaking, this formula significantly improves the
classical inclusion-exclusion sieve for distinct coordinate counting
problems. We cite it here without proof, and there are many
interesting applications of this new sieve
method \cite{lw,lw1,lwz}.\\

Let $\mathcal{P}$ be a finite set, and $\mathcal{P}^t$ denotes the
Cartesian product of $t$ copies of $\mathcal{P}.$ Let $X$  be a
subset of $\mathcal{P}^t.$ Define $\bar{X}=\{x=(x_1, x_2, \cdots,
x_t)\in X|x_i\neq x_j, i\neq j\},$ Let $f(x_1, x_2, \cdots, x_t)$ be
a complex valued function defined over $X$ and
$$F=\sum_{x\in \bar{X}}f(x_1, x_2, \cdots, x_t).$$

Let $S_t$ be the symmetric group on $\{1, 2, \cdots, t\}.$ Each
permutation $\tau\in S_t$ factorizes uniquely as a product of
disjoint cycles and each fixed point is viewed as a trivial cycle of
length $1.$ Two permutations in $S_t$ are conjugate if and only if
they have the same type of cycle structure (up to the order). For
$\tau\in S_t$, define the sign of $\tau$ to be
$\mathrm{sign}(\tau)=(-1)^{(t-l(\tau))},$  where $l(\tau)$ is the
number of cycles of $\tau$ including the trivial cycles. For a
permutation $\tau=(i_1 i_2 \cdots i_{a_1})(j_1 j_2 \cdots
j_{a_2})\cdots (l_1 l_2 \cdots l_{a_s})$ with $1\leq a_i, 1 \leq
i\leq s,$ define
$$X_{\tau}=\{(x_1, \cdots, x_t)\in X, x_{i_1}=\cdots=x_{i_{a_1}},\cdots,
x_{l_1}=\cdots=x_{l_{a_s}}\}.$$ For $\tau\in S_t,$ define
$F_{\tau}=\sum_{x\in X_{\tau}}f(x_1, x_2, \cdots, x_t).$ Now we can
state Li-Wan's sieve formula.

\begin{Theorem}
	Let $F$ and $F_{\tau}$ be defined as above. Then
	\begin{equation}\label{sum1}
	F=\sum_{\tau\in S_t}\mathrm{sign}(\tau)F_{\tau}.
	\end{equation}
\end{Theorem}

Note that the symmetric group $S_t$ acts on $\mathcal{P}^t$ naturally
by permuting coordinates, That is,for $\tau\in S_t$ and $x=(x_1, x_2
\cdots, x_t) \in \mathcal{P}^t,$ we have $ \tau\circ x=(x_{\tau(1)},
x_{\tau(2)}, \cdots, x_{\tau(t)}).$ A subset $X\subset \mathcal{P}^t$
is said to be symmetric if for any $x\in X$ and any $\tau\in S_t, \tau\circ
x \in X.$

For $\tau\in S_t,$ denote by $\bar{\tau}$ the conjugacy class
determined by $\tau$ and it can also be viewed as the set of
permutations conjugate to $\tau.$ Conversely, for one given conjugacy
class $\bar{\tau}\in C_t,$ denote by $\tau$ a representative
permutation of this class.  For convenience we usually identify
these two symbols.

In particular, if $X$ is symmetric and $f$ is a symmetric function
under the action of $S_t,$ we then have the following simpler
formula than (\ref{sum1}).

\begin{Corollary}
	Let $C_t$ be the set of conjugacy classes of $S_t.$ If $X$  is
	symmetric and $f$ is  symmetric, then
	\begin{equation}\label{sum2}
	F=\sum_{\tau\in C_t}\mathrm{sign}(\tau)C(\tau)F_{\tau},
	\end{equation}
	where $C(\tau)$ is the number of permutations conjugate to $\tau.$
\end{Corollary}

For the purpose of evaluating the above summation,  we need a
combinatorial formulas. A permutation $\tau \in S_t$ is said to be
of type $(c_1, c_2, \cdots, c_t)$ if $\tau$ has  exactly $c_i$
cycles of length $i.$ Note that $\sum_{i=1}^t ic_i=t.$ As we know,
two permutations in $S_t$ are conjugate if and only if they have the
same type of cycle structure. Let $N(c_1, c_2, \cdots, c_t)$ be the
number of permutations in $S_t$ of type $(c_1, c_2, \cdots, c_t)$
and  it is well known that
$$N(c_1, c_2, \cdots, c_t)=\frac{t!}{1^{c_1}c_1!2^{c_2}c_2!
	\cdots t^{c_t}c_t!}.$$

\begin{Lemma}\label{com}
	Define the generating function
	$$C_t(q_1, q_2, \cdots, q_t)=\sum_{\sum_{i=1}^t ic_i=t}N(c_1, c_2, \cdots, c_t)
	q_1^{c_1}q_2^{c_2}\cdots q_t^{c_t},$$  and set $q_1=q_2=\cdots=q_k=q$, then we have
	\begin{eqnarray*}
		C_t(q, q, \cdots, q)&=&\sum_{\sum ic_i=t}N(c_1, c_2, \cdots,
		c_t)q^{c_1}
		q^{c_2}\cdots q^{c_t}\\
		&=&(q+t-1)_t.
	\end{eqnarray*}
	If we set $q_i=q$ for $d\mid i$ and $q_i=s$ for $d\nmid
	i$, then
	\begin{equation*}
	\begin{array}{rl}
	&C_t(\overbrace{s,\cdots,s}^{d-1},q,\overbrace{s,\cdots,s}^{d-1},q,
	\cdots)\\
	=& \sum_{\sum
		ic_i=t} N(c_1,c_2,\cdots,c_t)s^{c_1}s^{c_2}\cdots q^{c_d}s^{c_{d+1}}\cdots\nonumber \\
	=& t!\sum_{i=0}^{\lfloor t/d \rfloor}{\frac{q-s}{d}+i-1\choose
		\frac{q-s}{d}-1} {s+t-di-1\choose s-1}\\
	\leq& t!{s+t+(q-s)/d-1\choose t}.
	\end{array}
	\end{equation*}
\end{Lemma}

Let $(\mathfrak{G},\oplus)$ be a finite Abelian group with order
$\big|\mathfrak{G}\big|=N,$ and let $\mathcal{P}\subset
\mathfrak{G}$ be a nonempty subset of cardinality $n.$
${\mathcal{P}\choose t}$ denotes the set of all t-subsets of
$\mathcal{P},$ then $\big|{\mathcal{P}\choose t}\big|={n \choose
	t}.$ For $B\in \mathfrak{G},$ let $$\mathfrak{N}(t, B,
\mathcal{P})=\left\{\mathcal{A}\in {\mathcal{P}\choose t}
\bigg|\mathop{\oplus}\limits_{P\in \mathcal{A}}P=B \right\},$$ and
$$N(t,B,\mathcal{P})=\big|\mathfrak{N}(t, B, \mathcal{P})\big|.$$
$\hat{\mathfrak{G}}$ is the group of additive characters of
$\mathfrak{G}$ with trivial character $\chi_0.$ Note that
$\hat{\mathfrak{G}}$ is isomorphic to $\mathfrak{G}.$ Denote the
partial character sum $s_{\chi}(\mathcal{P}) =\sum_{a\in
	\mathcal{P}}\chi(a)$ and the amplitude $\Phi(\mathcal{P})=
\mathop{\max}\limits_{\chi\in \hat{\mathfrak{G}},\chi\neq \chi_0}
|s_{\chi}(\mathcal{P})|.$

For our application, the estimate $N(t,B,\mathcal{P})$ is crucial,
when $\mathfrak{G}=\mathrm{Jac}(C)(\mathbb{F}_{q})$, which is the
rational points of the Jacobian variety of an algebraic curve. In
\cite{lw,lw1,lwz} the authors gave estimates for some special
finite Abelian groups. We use their method to give an estimate for a
general finite Abelian group.

\begin{Lemma}\label{bd}
	Let $N(t,B,\mathcal{P})$ be defined as above. Then
	\begin{equation}\label{sb}
	\left | N(t, B, \mathcal{P})-\frac{{n \choose t}} {N}\right| \leq{M \choose t},
	\end{equation}
	where $M$ is defined as $M=\max\left \{ \Phi(\mathcal{P})+t-1, \frac{n+\Phi(\mathcal{P})}{2}, \frac {n-\Phi(\mathcal{P})}{3}+\Phi(\mathcal{P})+t-1
	\right\}.$
\end{Lemma}
\begin{Proof}
	Let $X=\mathcal{P}^t$ be the Cartesian product of $t$ copies of
	$\mathcal{P},$ and $\bar X=\{x=(x_1, x_2 \cdots, x_t)\in
	\mathcal{P}^t|x_i\neq x_j, i\neq j\}.$ It is clear that
	$\big|X\big|=n^t$ and $\big|\bar X\big|=(n)_t.$ Then
	\begin{eqnarray*}
		t!N(t,B,\mathcal{P})&=& N^{-1}\sum_{x\in \bar{X}}\sum_{\chi\in
			\hat{\mathfrak{G}}}
		\chi(x_1+x_2+\cdot\cdot\cdot+x_t-B)\\
		&=&\frac{(n)_t}{N}+N^{-1}\sum_{\chi\neq \chi_0}\sum_{x\in \bar{X}}
		\chi(x_1)\chi(x_2)\cdots\chi(x_t)\chi^{-1}(B)\\
		&=&\frac{(n)_t}{N}+N^{-1}\sum_{\chi\neq \chi_0}\chi^{-1}(B)
		\sum_{x\in \bar{X}}\prod_{i=1}^t\chi(x_i).\\
	\end{eqnarray*}
	For $\chi\neq \chi_0,$ let $f_{\chi}(x)=f_{\chi}(x_1, x_2 \cdots,
	x_t)= \prod_{i=1}^t\chi(x_i),$ and for $\tau\in S_t$ let
	$$F_{\tau}(\chi)=\sum_{x\in X_{\tau}}f_{\chi}(x)=\sum_{x\in X_{\tau}}
	\prod_{i=1}^t\chi(x_i).$$ Obviously $X$ and $f_{\chi}(x_1, x_2
	\cdots, x_t)$ are symmetric. Applying (\ref{sum2}),
	$$t!N(t,B,\mathcal{P})=\frac{(n)_t}{N}+N^{-1}\sum_{\chi\neq \chi_0}
	\chi^{-1}(B) \sum_{\tau\in
		C_t}\mathrm{sign}(\tau)C(\tau)F_{\tau}(\chi).$$
	
	Assume $\tau$ is of type $(c_1, c_2, \cdots, c_t),$ without loss of
	generality, we can write
	$$\tau=(1)(2)\cdots(c_1)((c_1+1)(c_1+2))\cdots((c_1+2c_2-1)(c_1+2c_2))\cdots.$$
	One can check that $$X_{\tau}=\left\{(x_1,\cdots,
	x_t)\in\mathcal{P}^t\big|
	x_{c_1+1}=x_{c_1+2},\cdots,x_{c_1+2c_2-1}=x_{c_1+2c_2},\cdots\right\}.$$
	Then we have
	\begin{eqnarray*}
		F_{\tau}(\chi)&=& \sum_{x\in X_{\tau}}\prod_{i=1}^t\chi(x_i)\\
		&=&\sum_{x\in X_{\tau}}\prod_{i=1}^{c_1}\chi(x_i)\prod_{i=1}^{c_2}
		\chi^2(x_{c_1+2i})\cdots \prod_{i=1}^{c_t}\chi^t(x_{c_1+2c_2+
			\cdots ti})\\
		&=&\prod_{i=1}^{t}(\sum_{a\in \mathcal{P}}\chi^i(a))^{c_i}\\
		&=& n^{\sum c_i m_i(\chi)} s_{\chi}(\mathcal{P})^{{\sum c_i (1-m_i(\chi))}},
	\end{eqnarray*}
	where $m_i(\chi)=1$ if $\chi^i=1$ and otherwise $m_i(\chi)=0$.
	
	Now suppose $\mathrm{ord}(\chi)=d$ with $d\mid N$. Note that
	$C(\tau)=N(c_1,c_2,\dots,c_t)$. In the case $3\leq d\leq t$ since $|s_\chi(\mathcal{P})|\leq \Phi(\mathcal{P})$, applying Lemma \ref{com}, we
	have
	\begin{align*}
	&\sum_{\tau\in C_{t}}\mathrm{sign}(\tau)C(\tau) F_{\tau}(\chi)\\
	&\leq\sum_{\tau\in C_{t}}C(\tau) n^{\sum c_i m_i(\chi)}
	\Phi(\mathcal{P})^{{\sum c_i (1-m_i(\chi))}}\\
	&\leq t!{\frac {n-\Phi(\mathcal{P})}{d}+\Phi(\mathcal{P})+t-1 \choose t}.
	\end{align*}
	In the case $d=2$, it can also be prooved that \cite{lwz}
	$$\sum_{\tau\in C_{t}}\mathrm{sign}(\tau)C(\tau) F_{\tau}(\chi)\leq t!{\frac {n+\Phi(\mathcal{P})} 2\choose t}.$$
	
	Similarly, if $\mathrm{ord}(\chi)>t$,  then
	\begin{align*}
	\sum_{\tau\in C_{t}}\mathrm{sign}(\tau)C(\tau) F_{\tau}(\chi)\leq t!{
		\Phi(\mathcal{P})+t-1 \choose t}.
	\end{align*}
	
	Let $T$ be the set of characters which have order greater than $t$.
	Summing over all nontrivial characters,  we obtain
	\begin{align*}
	\left | N(t, B, \mathcal{P})-\frac{{n \choose t}} {N}\right| &\leq \frac {|T|}
	{N}{\Phi(\mathcal{P})+t-1 \choose t}+\frac {\pi(2)} {N}{\frac {n+\Phi(\mathcal{P})}
		2\choose t}\nonumber \\ &+\frac {1} {N}\sum_{2<d\leq t }
	\pi(d){\frac {n-\Phi(\mathcal{P})}{d}+\Phi(\mathcal{P})+t-1 \choose t},
	\end{align*}
	where $\pi(d)$ is the number of characters in $\widehat{\mathfrak{G}}$ of order $d$. The sequence $\{\frac {n-\Phi(\mathcal{P})}{d}+\Phi(\mathcal{P})+t-1\}_{d>2}, $ is decreasing. So let $$M=\max\left \{ \Phi(\mathcal{P})+t-1, \frac{n+\Phi(\mathcal{P})}{2}, \frac {n-\Phi(\mathcal{P})}{3}+\Phi(\mathcal{P})+t-1
	\right\},$$ we have the inequality.
	
\end{Proof}

\subsection{Abel-Jacobi Map}

Let $C/\mathbb{F}_{q}$ be a smooth projective curve of
genus $g$ over the finite field $\mathbb{F}_{q}.$ The divisor class
group of $C$ is defined to be the quotient group
$\mathrm{Pic}(C)=\mathrm{Div}(C)/\mathrm{Prin}(C),$ where
$\mathrm{Prin}(C)$ is the subgroup consisting of all principal
divisors. For a divisor $D\in \mathrm{Div}(C),$ the corresponding
element in the factor group $\mathrm{Pic}(C)$ is denoted by $[D],$
the divisor class of $D.$

We have the degree zero divisor class subgroup
$\mathrm{Pic}^0(C)=\mathrm{Div}^0(C)/\mathrm{Prin}(C).$ Assume that
$C$ has a $\mathbb{F}_{q}$-rational point $Q,$ it is well known that
there exists an Abelian variety $\mathrm{Jac}(C)/\mathbb{F}_{q}$ of
dimension $g$ with the property that for every extension field
$k/\mathbb{F}_{q},$  there is a naturally isomorphism
$$\mathrm{Jac}(C)(k)\longrightarrow \mathrm{Pic}^0(C)(k).$$
Moreover, the so-called Abel-Jacobi map $\phi_1: C/\mathbb{F}_{q}
\rightarrow \mathrm{Jac}(C)/\mathbb{F}_{q}$ given by
$$\phi_1: \quad P \mapsto [P-Q]$$ is a morphism of algebraic
varieties over $\mathbb{F}_{q}.$

Let $\mathrm{Sym}^d C$ denote the $d$-th symmetric product of
$C,$ and let $\mathrm{Div}^d_+(C)$ denote the set of the effective
rational divisors of degree $d.$ Then $\mathrm{Sym}^d
C(\mathbb{F}_q)$ can be identified with $\mathrm{Div}^d_+(C),$ And
let $A_d\stackrel{\mathrm{def}}{=}|\mathrm{Div}^d_+(C)|=
|\mathrm{Sym}^dC(\mathbb{F}_q)|.$
We can further define the $d$-th Abel-Jacobi map $\phi_d:
\mathrm{Sym}^d C \rightarrow \mathrm{Jac}(C)$ by $$\phi_d: \quad
D_1+\cdots+D_h \mapsto [D_1+\cdots+D_h-dQ],$$ where $D_1, \cdots,
D_h$ are closed points of $C/\mathbb{F}_q,$ with
$\sum_{i=1}^h\mathrm{deg}(D_i)=d.$ Let $W_d=\phi_d(\mathrm{Sym}^d
C)$ denote
the image of $\phi_d$ in $\mathrm{Jac}(C).$

The Abel-Jacobi Theorem says that $\phi_1: C\rightarrow W_1$ is an
isomorphism, $\phi_d:\mathrm{Sym}^d C\rightarrow W_d$ are birational
morphisms, for $2\leq d\leq g,$ and $W_g=\mathrm{Jac}(C).$ Let
$h_q(C)=|\mathrm{Jac}(C)(\mathbb{F}_q)|,$ then there is Hasse-Weil
bound for $\mathrm{Jac}(C)$ \cite{nx,tv}
\begin{equation}\label{jb}
(\sqrt{q}-1)^{2g}\leq h_q(C) \leq (\sqrt{q}+1)^{2g}.
\end{equation}

We have the following estimates (For details please refer
to  \cite{nx}, Lemma 5.3.4) as well.

\begin{Proposition}
	Let $C/\mathbb{F}_q$ be an algebraic curve of genus $g\geq1.$ Then
	for any integers $d\geq0$ we have
	$$
	A_d\leq\frac{h_q(C)}{q^{g-d}}\left(\frac{2gq^{1/2}}{q^{1/2}-1}-
	\frac{q}{q-1}\right).
	$$
\end{Proposition}

Consequently, we have
\begin{equation}\label{wb}
\big|W_d(\mathbb{F}_q)\big|\leq
A_d\leq\frac{h_q(C)}{q^{g-d}}\left(\frac{2gq^{1/2}}{q^{1/2}-1}-
\frac{q}{q-1}\right),
\end{equation} for
$1\leq d\leq g-1.$

\subsection{Character Sums on Curve}

$\mathrm{Jac}(C)(\mathbb{F}_{q})$ is a finite Abelian group. Let
$\chi: \mathrm{Jac}(C)(\mathbb{F}_{q})\rightarrow \mathbb{C}^*$ be a
character. Via Abel-Jacobi map $\phi_1: \quad P \mapsto [P-Q],$ one
can consider character sum on curve $C$, let
$$s_\chi=\sum_{P\in C(\mathbb{F}_q)}\chi([P-Q]).$$
If $\chi$ is trivial, we have Hasse-Weil bound
$$\big|s_\chi-q\big|\leq 2gq^{\frac{1}{2}}.$$
If $\chi$ is nontrivial, we have the following Weil bounds for character sums (For details
please refer to \cite{katz}, proposition 9.1.3 and \cite{rosen},
chapter 9).

\begin{Proposition}\label{charS}
	Suppose $\chi$ is a nontrivial character of
	$\mathrm{Jac}(C)(\mathbb{F}_{q}),$ then
	\begin{equation}\label{dwb}
	|s_\chi|\leq (2g-2)
	q^{\frac{1}{2}}.
	\end{equation}
\end{Proposition}

\section{AGLSSS over large fields are asymptotically threshold}

In this section we consider the case of secret sharing schemes from
elliptic curves when $q \rightarrow \infty$ at first, since in
this case Theorem 1 can be used to count
qualified and unqualified sets directly.\\

Let $E/\mathbb{F}_{q}$ be an elliptic curve, and let
$C/\mathbb{F}_{q}$ be an algebraic curve of genus $g$ at least $2.$
In this section we let the Abelian group $\mathfrak{G}$ be equal to
$E(\mathbb{F}_{q})$ and $\mathrm{Jac}(C)(\mathbb{F}_{q})$
separately. $\oplus$ and $\ominus$ denote additive and minus
operator in the group.

\subsection{Elliptic curve case}
Let $\mathfrak{G}=E(\mathbb{F}_{q})$ with zero element $O.$ As in
Theorem \ref{CLX}, $\{P_0, \cdots, P_n\}$ are a subset of
$E(\mathbb{F}_q)$ of $n + 1$ nonzero distinct elements.
$D=P_0+\cdots +P_n$ and $G=mO$ are divisors of $E.$ We have the
secret sharing scheme from $C_{\Omega}(D, G),$ with the set of
players $\mathcal{P}=\{P_1, \cdots,  P_n\}.$  Then we have the following
result.

\begin{Theorem}\label{ESSS} Suppose that $m=\delta n,$ where $\delta$
	is a constant between $0$ and $\frac{2}{3}.$  As $q\rightarrow \infty,$ we
	assume that $n$ and $m$ all approach to infinity, and
	$\big|\mathfrak{G}\big|-\big|\mathcal{P}\big|$ is bounded by a constant
	$c.$ Then\\
	I) when $t=m,$ the proportion of the qualified subsets approaches to
	$zero$;\\
	II) when $t=m-1,$ the proportion of the qualified subsets approaches
	to 1.
\end{Theorem}
\begin{Proof}
	Note that $\mathcal{P}$ is the set of players $\{P_1,
	\cdots,  P_n\}.$
	Recall that ${\mathcal{P}\choose t}$ denotes the set of all $t$-subsets of
	$\PP.$
	
	Let $\mathcal{A}\in {\mathcal{P}\choose t}$ be a subset of $\PP$ with cardinality $t$. Then
	$\mathcal{A}^c =\mathcal{P}\setminus\mathcal{A}$ is a qualified
	subset if and only if we can find a rational function $f\in L(G)$ such that
	$f(P_0)\neq0$ and
	$f(P) = 0$ for all $P\in \A $.
	
	Now we divide into two sub-cases for completing the whole proof of the theorem.
	\begin{itemize}
		\item[I)]
		Suppose $t=m.$ Because $f$ is a nonzero element of $L(G)$ with the divisor $G=mO$,
		it has at most $m$ distinct zeros. It follows from the assumption $t=m$
		that the divisor of the function $f$ should be
		$$\mathrm{div}(f)=\sum_{P\in
			\mathcal{A}}P-mO.$$ The existence of such a function $f$ is equivalent to
		saying (see Theorem 11.2 of \cite{washington} )
		
		$$\mathop{\oplus}\limits_{P\in \mathcal{A}}P=O.$$
		Namely
		$\mathcal{A}\in \mathfrak{N} (t, O, \mathcal{P}).$
		For our purpose,  we now give an estimate for the value $\Phi(\PP)$.
		According to the basic properties of group characters, we have
		$$\sum_{g\in
			\mathfrak{G}}\chi(g)=0$$ for a nontrivial character $\chi$.
		This implies that
		$$\sum_{g\in
			\PP}\chi(g)= -\sum_{g\in \mathfrak{G}-\PP}(g) $$
		and so
		$$\Phi(\mathcal{P})=\Phi(\mathfrak{G}-\mathcal{P})\leq
		\big|\mathfrak{G}-\PP \big|=
		\big|\mathfrak{G}\big| -\big|\mathcal{P}\big|\leq c,$$
		if $\big|\mathfrak{G}\big|-\big|\mathcal{P}\big|$ is bounded by
		a constant $c.$ Now let $n\rightarrow \infty,$ if $t<\frac{1}{6}n,$ then $M=\frac{n+\Phi(\mathcal{P})}{2};$ if $\frac{1}{6}n \leq t=m<\frac{2}{3}n,$ then  $M=\frac {n-\Phi(\mathcal{P})}{3}+\Phi(\mathcal{P})+t-1.$ in a word, $M<\delta'n,$ where $0<\delta'<1.$ It follows from Lemma \ref{bd} that
		\begin{eqnarray*}
			\frac{N(t,O,\mathcal{P})}{{n \choose t}}&\leq& \frac{1}{N}+
			\frac{{M\choose t}}{{n \choose t}}\\
			&\leq& \frac{1}{N}+\frac{{\delta'n\choose t}}
			{{n \choose t}}\\
			&=& \frac{1}{N}+\frac{(\delta'n)(\delta'n-1)\cdots (\delta'n-t+1)}{n(n-1)\cdots(n-t+1)}.   \\
		\end{eqnarray*}
		Since $$1>\delta'=\frac{\delta'
			n}{n}>\frac{\delta'n-1}{n-1}> \cdots
		>\frac{\delta'n-t+1}{n-t+1},$$
		it follows that $\lim\limits_{q\rightarrow \infty}\frac{(\delta'n)(\delta'n-1)\cdots (\delta'n-t+1)}{n(n-1)\cdots(n-t+1)}=0.$ Note that $\lim\limits_{q\rightarrow \infty}\frac{1}{N}=0.$
		Hence we have
		$$\lim\limits_{q\rightarrow \infty}\frac{N(t,O,\mathcal{P})}{{n \choose t}}
		=0.$$ This means that
		the proportion of the qualified subsets approaches to
		$zero$ if $t=m$.
		
		\item[II)]  Suppose $t=m-1.$ For the same reason, the divisor of the function $f$ should satisfy
		$$\mathrm{div}(f)=\sum_{P\in \mathcal{A}}P+P'-mO,$$ where $P'\in
		E(\mathbb{F}_q)$ and $P'\neq P_0.$ Similarly, the existence of the function $f$ is equivalent to
		$$\mathop{\oplus}\limits_{P\in \mathcal{A}}P=\ominus P'\neq P_0.$$ That is
		to say,  $\mathcal{A}\notin\mathfrak{N}(t, (\ominus P_0),
		\mathcal{P}).$ Similarly, one can show that
		$$\lim\limits_{q\rightarrow \infty} \frac{N(t,(\ominus
			P_0),\mathcal{P})}{{n \choose t}}=0.$$ This implies that the proportion of the
		qualified subsets which is equal to $1-\frac{N(t,(\ominus
			P_0),\mathcal{P})} {{n \choose t}}$ approaches to $1$ if  $q$ and $m$ tend to infinity and
		$t=m-1$. This completes the whole proof of Theorem~\ref{ESSS}.
	\end{itemize}
\end{Proof}

\subsection{General curve case}
Let $C/\mathbb{F}_q$ be an algebraic curve of genus $g$. Let $\{Q,
P_0, P_1,\cdots,P_n\}$ be a subset of $C(\mathbb{F}_q)$, and let
$D=P_0+\cdots +P_n$ and $G=mQ.$ We have the secret sharing scheme
from $C_{\Omega}(D, G),$ with the set of players $\mathcal{P}=\{P_1,
\cdots, P_n\}.$ In this section, we will consider the asymptotic
access structures of that algebraic geometric secret sharing schemes
as $q\rightarrow \infty.$

We denote $\mathfrak{G}=\mathrm{Jac}(C)(\mathbb{F}_{q})$ in this
subsection.  Because Abel-Jacobi map $\phi_1:C\rightarrow
\mathrm{Jac}(C),$ is an embedding. For a subset $\mathcal{S}\subset
C(\mathbb{F}_q),$  the symbol $\mathcal{S}$ sometimes denotes its
image $\phi_1(\mathcal{S})\subset \mathrm{Jac}(C)(\mathbb{F}_{q})$
by abuse of notation.  By Hasse-Weil bound we have $q-2gq^{1/2}\leq
\big|C(\mathbb{F}_q)\big|\leq q+2gq^{1/2}.$ If the genus $g$
satisfies $\lim\limits_{q\rightarrow \infty}\frac{g}{\sqrt{q}}=0,$
then $\big|C(\mathbb{F}_q)\big|\sim q$.   Thus
$\big|C(\mathbb{F}_q)\big|$ and
$h_q(C)=\big|\mathrm{Jac}(C)(\mathbb{F}_q)\big|\geq
(q^{1/2}-1)^{2g}$ both approach to infinity as $q$ tends to
infinity.  We prove the following result.

\begin{Theorem}\label{AGSSS} We assume that $q \rightarrow \infty $ and
	$\lim\limits_{q\rightarrow \infty}\frac{g}{\sqrt{q}}=0.$ Suppose
	that $m=\delta n,$ where $\delta$ is a constant between $0$ and $\frac{2}{3},$
	$m$ and $n$ go to the infinity as $q$ tends to infinity. Suppose
	that $\big|C(\mathbb{F}_q)\big|-\big|\mathcal{P}\big|$ is bounded by
	a
	constant $c$ as $q$ tends to infinity.  Then
	\begin{itemize}
		\item[I)]
		when $0\leq m-t<g,$ the proportion of the qualified subsets
		approaches to zero;
		\item[II)]
		when $g\leq m-t<2g,$ the proportion of the qualified subsets
		approaches to~1.
	\end{itemize}
\end{Theorem}

\begin{Proof}
	For any $\mathcal{A}\in {\mathcal{P}\choose t},$ let
	$\mathcal{A}^c=\mathcal{P}\setminus \mathcal{A}$ and
	$P_{\mathcal{A}}=\sum_{P\in \mathcal{A}}P\in \mathrm{Div}^t_+(C).$
	Then $\mathcal{A}^c$ is a qualified subset if and only if we can find a
	rational function $f\in L(G)$ such that $\mathcal{A}$ is contained in the
	zero locus of $f$ and $f(P_0)\neq0,$ i.e. $f\in
	L(G-P_{\mathcal{A}})\setminus L(G-P_{\mathcal{A}}-P_0).$ And by the
	Riemann-Roch Theorem~\cite{stichtenoth}, we have the following equivalent condition for
	$\mathcal{A}^c$ qualified
	\[\left\{ \begin{array}{ll}
	l(G-P_{\mathcal{A}})>0, & (a)\\
	L(K_C-G+P_{\mathcal{A}})=L(K_C-G+P_{\mathcal{A}}+P_0), & (b)
	\end{array}\right. \]
	where $l(G-P_{\mathcal{A}})=\dim (L(G-P_{\mathcal{A}})),$ and $K_C$
	is the canonical divisor of $C.$
	
	\begin{itemize}
		\item[I)] Suppose $0\leq m-t< g.$ If $\mathcal{A}^c$ is qualified, then
		$$l(G-P_{\mathcal{A}})
		>0,$$
		by the condition $(a).$ We choose a nonzero
		element $f\in L(G-P_{\mathcal{A}}),$ and let
		$D_{\mathcal{A}}=G-P_{\mathcal{A}}+\mathrm{div}(f)\geq0$ which is an
		effective divisor of $C,$ with $\mathrm{deg}(D_{\mathcal{A}})=m-t.$
		
		Consider the following map ${\mathcal{P}\choose t}\hookrightarrow
		\mathrm{Div}^t_+(C)\rightarrow \mathrm{Jac}(C)$ given by
		$$\mathcal{A}\mapsto P_{\mathcal{A}} \stackrel{\phi_t}{\mapsto}
		[P_{\mathcal{A}}-tQ]. $$
		
		Because $D_{\mathcal{A}}$ is linearly
		equivalent to $G-P_{\mathcal{A}},$ we have the divisor class equality
		$$[D_{\mathcal{A}}]=[G-P_{\mathcal{A}}]$$ in the Jacobian group
		$\mathrm{Jac}(C)$. It follows from $G= mQ$ that
		$$[D_{\mathcal{A}}-(m-t)Q]=[G-P_{\mathcal{A}}-(m-t)Q]
		=\ominus[P_{\mathcal{A}}-tQ].$$ Thus
		$$\phi_t(P_{\mathcal{A}})=\ominus\phi_{m-t}(D_{\mathcal{A}})\in \ominus
		W_{m-t},$$
		where $\ominus
		W_{m-t}$ denotes the set that contains all negative elements in $W_{m-t}$.
		
		Now we get $$ \mathcal{A} \in \bigcup_{a\in \ominus
			W_{m-t}}\phi_t^{-1}(a)\cap
		{\mathcal{P}\choose t}=\bigcup_{a\in \ominus W_{m-t}}\mathfrak{N}
		(t, a, \mathcal{P})
		$$
		by the definition of $\mathfrak{N}
		(t, a, \mathcal{P})$.
		
		For our purpose, we give an estimate for $\Phi(\mathcal{P})$.
		By Proposition~\ref{charS}, for a
		nontrivial character $\chi,$ we have
		$$\Phi(\mathcal{P})=\Phi(C(\mathbb{F}_q)-(C(\mathbb{F}_q)\backslash
		\mathcal{P}))\leq\Phi(C(\mathbb{F}_q))+\Phi(C(\mathbb{F}_q)\backslash
		\mathcal{P}))\leq (2g-2)q^{\frac{1}{2}}+c$$ if
		$\big|C(\mathbb{F}_q)\big|-\big|\mathcal{P}\big|$ is bounded by a constant $c.$ Since $\lim\limits_{q\rightarrow \infty}\frac{(2g-2)\sqrt{q}}{n}=0.$ In Lemma \ref{bd}, as the elliptic curve case, we have $M<\delta'n,$ where $\delta'<1.$
		
		We are now in a position to show that  the proportion of the qualified
		subsets tends to $zero$ , i.e., the ratio
		
		$$  {\sum\limits_{a\in \ominus W_{m-t}}N(t,a,\mathcal{P})}\Big/{{n
				\choose t}} \rightarrow 0,$$
		as $q$ tends to infinity

		On basis of (\ref{sb}), (\ref{jb}) and (\ref{wb}) one has
		\begin{eqnarray*}
			& &\frac{\sum\limits_{a\in \ominus W_{m-t}}N(t,a,\mathcal{P})}{{n
					\choose t}}\\
			&\leq& \frac{\big|\ominus W_{m-t}(\mathbb{F}_q)\big|}{{n
					\choose t}}\left\{\frac{{n \choose
					t}}{h_q(C)}+{M
				\choose t}\right\} \\
			&<& \frac{\big|W_{m-t}(\mathbb{F}_q)\big|}{h_q(C)}+\big|W_{m-t}(\mathbb{F}_q)\big|
			\frac{(\delta'n)_t}
			{(n)_t}\\
			&\leq & \frac{1}{q^{(g-(m-t))}}\left(\frac{2gq^{1/2}}{q^{1/2}-1}-
			\frac{q}{q-1}\right)+\\
			& &\frac{h_q(C)}{q^{(g-(m-t))}}\left(\frac{2gq^{1/2}}{q^{1/2}-1}-
			\frac{q}{q-1}\right)\prod_{i=0}^{t-1}\frac{\delta'n-i}{n-i}\\
			&\leq & \frac{1}{q}\left(\frac{2gq^{1/2}}{q^{1/2}-1}-
			\frac{q}{q-1}\right)+\\
			& &\frac{(q^{1/2}+1)^{2g}}{q}\left(\frac{2gq^{1/2}}{q^{1/2}-1}-
			\frac{q}{q-1}\right)\prod_{i=0}^{t-1}\frac{\delta'n-i}{n-i},\ \ ( \star)\\
		\end{eqnarray*}
		where $n\sim q$ and $t\sim \delta q.$
		
		When $q\rightarrow \infty,$ we have
		$\frac{1}{q}\left(\frac{2gq^{1/2}}{q^{1/2}-1}-
		\frac{q}{q-1}\right)\rightarrow 0,$  since
		$\lim\limits_{q\rightarrow \infty}\frac{g}{\sqrt{q}}=0.$ And $\frac{\delta'n-i}{n-i}<\delta'<1,$ this implies that
		$$\prod_{i=0}^{t-1}\frac{\delta'n-i}{n-i}\leq
		\delta'^{\delta q}.$$ Consequently
		\begin{eqnarray*}
			&\ln &\left\{\frac{(q^{1/2}+1)^{2g}}{q}
			\left(\frac{2gq^{1/2}}{q^{1/2}-1}-
			\frac{q}{q-1}\right)\prod_{i=0}^{t-1}\frac{\delta'n-i}{n-i}\right\}\\
			&\leq& 2g\ln(q^{1/2}+1)-\ln q+\ln\left(\frac{2gq^{1/2}}{q^{1/2}-1}-
			\frac{q}{q-1}\right)+(\delta\ln \delta')q\\
			&\rightarrow& -\infty,\\
		\end{eqnarray*}
		as $q$ tends to infinity.
		Thus $( \star)$ approaches to $zero$. The conclusion follows.\\
		
		\item[II)] Suppose $g\leq m-t< 2g.$ By the Riemann-Roch Theorem, we have
		$$l(G-P_{\mathcal{A}})\geq \deg(G-P_{\mathcal{A}})+1-g= m-t+1-g >0.$$
		This means that Condition $(a)$ always holds. It follows from
		the Condition $(b)$ that the set $\mathcal{A}^c$ is
		unqualified if and only if the set $L(K_C-G+P_{\mathcal{A}})$ is a proper
		subset of $L(K_C-G+P_{\mathcal{A}}+P_0)$, i.e.,  $$L(K_C-G+P_{\mathcal{A}})
		\subsetneq L(K_C-G+P_{\mathcal{A}}+P_0).$$  Let us
		choose a nonzero element $f \in
		L(K_C-G+P_{\mathcal{A}}+P_0)\setminus L(K_C-G+P_{\mathcal{A}}),$ and
		let
		$D_{\mathcal{A}}^K=K_C-G+P_{\mathcal{A}}+P_0+\mathrm{div}(f)\geq0$
		which is an effective divisor of $C,$ with degree
		$s=\mathrm{deg}(D_{\mathcal{A}}^K)=2g-1-(m-t),$ and $0\leq s\leq
		g-1.$ This fact will play a vital role in the proof.
		
		In a similar manner, we consider the following map ${\mathcal{P}\choose t}\hookrightarrow
		\mathrm{Div}^t_+(C)\rightarrow \mathrm{Jac}(C)$ given by
		$$\mathcal{A}\mapsto P_{\mathcal{A}} \stackrel{\phi_t}{\mapsto}
		[P_{\mathcal{A}}-tQ].$$ Because $D_{\mathcal{A}}^K$ is linearly
		equivalent to $K_C-G+P_{\mathcal{A}}+P_0,$ by definition,
		$[D_{\mathcal{A}}^K]=[K_C-G+P_{\mathcal{A}}+P_0],$ then
		\begin{eqnarray*}
			[D_{\mathcal{A}}^K-sQ]&=&[K_C-G+P_{\mathcal{A}}+P_0-sQ]\\
			&=& [K_C+P_0-(2g-1)Q]\oplus[P_{\mathcal{A}}-tQ].
		\end{eqnarray*} Thus
		\begin{eqnarray*}
			\phi_t(P_{\mathcal{A}})&=&\phi_s(D_{\mathcal{A}}^K)
			\ominus[K_C+P_0-(2g-1)Q]\\
			&\in & \phi_s(C_s)\ominus[K_C+P_0-(2g-1)Q]=\widetilde{W}_s,
		\end{eqnarray*}
		where $\widetilde{W}_s$ is the translation of $W_s$ by the element
		$\ominus[K_C+P_0-(2g-1)Q].$
		
		Namely, $\mathcal{A}^c$ is a qualified subset if and only if
		$$
		\mathcal{A} \notin \bigcup_{a\in \widetilde{W}_s}\phi_t^{-1}(a)\cap
		{\mathcal{P}\choose t}=\bigcup_{a\in \widetilde{W}_s}\mathfrak{N}
		(t, a, \mathcal{P}).
		$$ Since $s \leq g-1$, similarly one can show that
		$$\lim\limits_{q\rightarrow \infty}
		\frac{\sum_{a\in \widetilde{W}_s}N(t,a,\mathcal{P})}{{n \choose
				t}}=0.$$ So the proportion of the qualified subsets approaches to
		$1$ as $q$ tends to infinity.  This completes the whole proof of the theorem.
	\end{itemize}

\end{Proof}

\section{Conclusion}
Algebraic geometric secret sharing schemes have been widely used in
two-party secure computation, secure multiparty computation,
communication-efficient zero-knowledge of circuit satisfiability,
correlation extractors and OT-combiners since the publication of
\cite{cc}. In many cases when Shamir's threshold secret sharing
scheme is replaced by algebraic geometric secret sharing schemes the
base field can be constant-size and communication complexity can be
saved by a $log n$ factor. Thus it is quite important to answer the
question how far from threshold these quasi-threshold algebraic
geometric secret sharing schemes are. We showed that when the size
$q$ of the base field goes to the infinity and $\lim
\frac{g}{\sqrt{q}}=0$, algebraic geometric secret sharing schemes
are asymptotically threshold. It would be interesting to know
asymptotic situation of algebraic geometric secret sharing schemes
in the case that the size $q$ of base field is fixed and the genus
of curves goes to the infinity. In particular are algebraic
geometric secret sharing schemes asymptotically threshold in this
case?

\end{document}